# System Fitness and the Extinction Patterns of Firms under Pure Economic Competition

Paul Ormerod ( pormerod@volterra.co.uk )*

Helen Johns ( hjohns@volterra.co.uk )

Volterra Consulting Ltd

The Old Power Station
121  Mortlake High Street
London SW14 8SN

September 2001

\* Corresponding author

# Abstract

*The relationships which characterise the outcomes of the interactions between firms in the economy appear to follow power law behaviour. In particular, there is evidence that the empirical power laws which relate the size of an extinction to its frequency are similar for both biological species and firms in the economy.*

*Biological species interact with each other in ways which either increase or decrease the fitness levels of individual species. In the same way, firms in the economy carry out strategies, and the net outcome of the interactions between firms is to increase or decrease the fitness of other firms..*

*Models which are used to describe the interactions of biological species can be given straightforward interpretations in terms of the interactions of firms in an industry. The general versions of models of interactions between agents allow for co-operative behaviour between pairs of agents. We contrast the properties of such a model with special cases of it, in which interactions are restricted so that only competitive interactions between pairs of agents are permitted. Such pure economic competition (PEC) behaviour is regarded as desirable by economic theory,*

*The properties of the two models are strikingly different. A power law no longer provides a good description of the relationship between the size and frequency of extinctions in the PEC model, in which there are substantial numbers of very large extinctions. As a result, the level of fitness of the system as a whole remains considerably lower than in the general version of the model. This suggests that models of pure economic competition do not provide a good description of how many industries actually behave, and that a certain amount of collaboration and co-operation between firms is required for the survival of an industry.*

## 1. Introduction

The relationships which characterise the outcomes of the interactions between firms in the economy appear to follow power law behaviour. Economic recessions occur when, as a result of the output decisions of individual firms, total output growth falls below zero. A power law with exponent -1.7 gives a reasonable description of the duration of recessions in 17 Western economies over the period 1871-1994 [1]. An analysis of the extinction rates of the world's 100 largest industrial companies in 1912 over the period from 1912 to 1995 shows that a power law with estimated exponent of close to -2 gives a good empirical description of the data [2].

The size distribution of extinction events amongst biological species can also be described by a power law with an exponent of around -2 [3-5]. A number of models based upon the principle of self-organised criticality of a system arising from interactions between agents have been developed to account for this empirical power law relationship (for example, [6 - 9], with a recent survey given by [10] )[1].

The models which are used to describe the interactions of biological species can be interpreted in terms of the interactions of firms in an economy.

An important implication of economic theory is that it is desirable for firms in any given industry to compete with one another. A world in which pure competition of this kind exists implies restrictions on the ways in which agents are allowed to interact. In this paper, we contrast the properties of a general version of a model of species or agent interaction with that of a 'pure economic competition' version, which contains the required restrictions in the model. The model used is that of Solé and Manrubia [6], referred to subsequently as S-M, which has a particularly natural interpretation in an economic context. (A similar approach, developed quite independently, is used to describe the observed distribution of long-run growth rates across the Western capitalist economies [12]).

---

[1] A potential alternative approach to that of self-organised criticality to account for observed power law behaviour in the social sciences is given in [11]

We demonstrate that systems in which the restrictions on interactions between agents implied by economic theory do not in general give rise to power law behaviour in terms of the extinction patterns of agents. Further, the overall level of industry fitness which emerges in a purely competitive model is low, implying that growth in such systems is not easy to achieve. Co-operative (or collusive) interactions are required in order to generate power law behaviour of agent extinction, and such interactions lead to distinctly higher levels of industry fitness.

The findings also have policy implications in terms of the regulation of industries. Collusion between firms, of whatever form, is held to reduce the overall level of social welfare (see almost any economics textbook e.g.[13]). Most economic regulatory bodies are set up on exactly this principle: to promote competition between firms. Yet this has an adverse impact on the overall fitness of the industry.

Section 2 describes the S-M model, the economic interpretation of its principles, and how these are reflected in a pure economic competition version of the model. Section 3 sets out the properties of the S-M model populated by limited numbers of agents such as are often found in individual industries at any point in time. Section 4 contrasts these with the properties of the pure economic competition (PEC) version. Section 5 provides a conclusion.

## 2. The Solé - Manrubia model and its economic interpretation

The model contains N species and a matrix of couplings, $J_{ij}$, which indicates how each species i affects every other species j, with $J_{ij} \in [-1, 1]$. The model is solved over a sequence of iterated steps, and at each iteration the following occurs: i) for each species i, one of its $J_{ij}$ is replaced with a new value chosen at random from a uniform distribution on [-1, 1] ii) the overall fitness of any given species is measured by $f_i = \sum_j J_{ij}$, and any species for which $f_i < 0$ is deemed to be extinct. If m species become extinct, an avalanche of size m is deemed to have taken place iii) an extinct species is replaced by a new entrant into the system, which is very similar to that of one of the surviving species. Specifically, a surviving species k is chosen at random to replace each extinct species j, and the linkages $J_{ij}$ and $J_{ji}$ are replaced with $J_{ik} + \eta$ and $J_{ki} + \eta$, where $\eta$ is chosen at random from a small interval $[-\varepsilon, +\varepsilon]$.

The biological interpretation of the model is set out in [6, 10]. In an economic context, the $J_{ij}$ matrix can be thought of as the way in which the net impacts of their overall strategies lead individual agents, or firms, to interact in an economy or industry at any point in time. Three combinations of pair-wise connections are possible in terms of the signs of the $J_{ij}$: i) $J_{ij}, J_{ji} > 0$; ii) $J_{ij} > 0, J_{ji} < 0$, or vice versa; and iii) $J_{ij}, J_{ji} < 0$

Case (i) represents a situation in which firms benefit from each other's presence in a market. This could arise by collusive behaviour when the firms deliberately decide not to compete. The situation could also arise when, for example, two firms are independently opening up a new niche in a market. Marketing activity such as advertising by each firm creates greater awareness of the new kind of product, from which the brands of both firms can benefit.

Case (ii) arises when two firms are in competition, and the overall strategy of one firm is such that it gains fitness at the expense of its rival. Case (iii) is a more intense example of the competitive case (ii). In this instance, the degree of competition is such that the firms carry out actions which reduce both their fitness levels. An example is when two firms become engaged in a price war which ultimately reduces both their profit levels (other illustrations of such behaviour are discussed in, for example, [14]). Such a situation is unlikely to persist for any length of time, because it increases the chances of both firms becoming extinct.

Economic theory recognises that under conditions of uncertainty it is impossible for individual agents to explicitly follow maximising behaviour with respect to their fitness, because no one knows with certainty the outcome of a decision. Maximisation nevertheless occurs, because competition dictates that the more efficient firm will survive and the inefficient ones perish (the classic statement of this is [15]). This can be expressed as the updating rule for the $J_{ij}$ in the model, in which each firm develops its strategy by a process of trial-and-error.

The rule in the S-M model that extinct species are immediately replaced by new entrants implies that the economic interpretation has most relevance in industries in which new entry is relatively easy. This appears to be the case across a substantial part of the economy as a whole. For example, it is almost always the case in new industries [16]. Even in the US car industry, in its early stages during the first two decades of the twentieth century, no fewer than 1,641 producers participated at some time in the market [17]. Entry may be facilitated into more mature industries by regulatory change, (e.g. airlines and energy supply), by rapid technological innovation (e.g. the undermining of IBM by the development of the PC), or by a combination of both (e.g. telecommunications, financial services).

## 3. Properties of the S-M model

The purpose of this section is to describe briefly some key properties to enable comparisons with the pure economic competition version of the model to be made. We discuss the relationship between the size of extinction events and their frequency, and the average fitness of the system.

The size distribution of extinction events in the model is reported to follow a power law with exponent of around -2 [6], based upon versions of the model populated by 100 or 150 agents. The value of the exponent appears to depend upon the number of agents in the model, with results for 100, 200, 500 and 1000 agents shown in [10]. The dependence of the exponent on system size can be seen very clearly when smaller numbers of agents are used, which is probably more realistic in the context of single industry rather than in the general setting of biological species extinction. For $N = 25$, the least squares fitted estimate of the exponent of a power law, averaged over 500 solutions of the model each solved over 50,000 iterations and the first 10,000 omitted to eliminate transient behaviour from initial conditions (as in [6]), is -3.36; for N =50 it is -2.92, and for N = 100 it is -2.43. In each case, a power law does provide a reasonable description of the data. The summary statistics for N = 100 are as follows:

| Minimum | 1st Quartile | Median | Mean | 3rd Quartile | Maximum |
|---------|--------------|--------|-------|--------------|---------|
| 2.239   | 2.383        | 2.427  | 2.427 | 2.469        | 2.599   |

The overall level of fitness of the system is of interest both in a biological and an economic context. The maximum fitness which the model can take is theoretically $N^2$. In practice, the system does not approach this level, but the average value of fitness over the iterations of each solution of the model is greater than zero (this average value varies very little between solutions of the model containing the same number of agents).

The overall fitness of the system, as a proportion of the theoretical maximum for a given number of agents, rises slightly as the number of agents falls. This is implied by analytical results on the probability of the sign of the connection between any two agents being positive or negative [18], and the prediction is confirmed empirically. The overall fitness of the system is averaged in each solution from iteration 10,001 to 50,000, and the mean of these values over 500 separate solutions is taken. This figure is divided by the theoretical maximum for the number of agents, to give average fitness figures of 0.167 for N = 25, 0.143 for N = 50, and 0.118 for N = 100.

## 4. Properties of the Pure Economic Competition model

The key feature of this model is that firms cannot benefit from each other's activities ( connections of the form $J_{ij}$, $J_{ji} > 0$ are excluded). In a practical context, this could be brought about by regulation supported, if required, by legal sanctions.

We report here the results of the PEC model when the interactions between pairs of agents are all of the form $J_{ij} > 0$, $J_{ji} < 0$ ( or vice versa ), which is the usual form of competition between firms. Allowing pair-wise connections of the form $J_{ij}$, $J_{ji} < 0$ leads to the properties of the model being even more different from those of the general version, with complete extinction of the entire set of agents being a not infrequent event. Clearly, if such an interaction holds for any length of time, the probability of both firms becoming extinct is high.

Two versions of the PEC model are examined. In version A, we retain the rules of the S-M model for the initialisation of the J matrix and for the update rule for each agent at each iteration. The replacement rule is as follows. When offspring species are being created, the replacement $J_{ij}$ are at first drawn in accordance with the normal Sole-Manrubia rules. They are then checked for their sign relationship. If $J_{ij} > 0$ and $J_{ji} > 0$, then the sign of one of them switches sign with probability 1- $P_{+ve}$, where $P_{+ve}$ is the probability that a positive-positive relationship will be allowed to be remain. Similarly, if $J_{ij} < 0$ and $J_{ji} < 0$, one of them switches sign with probability 1- $P_{-ve}$. The general Sole-Manrubia model obtains when $P_{+ve} = P_{-ve} = 1$. In version A of the PEC model, we report results when $P_{+ve} = P_{-ve} = 0$ i.e. the sign of offspring connections is changed at random[2].

In version B of the PEC model, both the initialisation of the J matrix and the update rule are subject to the condition that if the signs of $J_{ij}$ and $J_{ji}$ match, one of them is switched at random. The opposite-signs-only condition is maintained in the replacement rule by ensuring that the new $J_{ij}$ granted to new entrant preserve the signs of the $J_{kj}$ on which the new entrant is based. In the general model, the new $J_{ij}$ takes the value $J_{kj} + \eta$, where k is the species chosen at random as the agent to be imitated and where $\eta$ is a small random number drawn from [-ε, ε]. In this version of the PEC model, if this leads to $J_{ij}$ being a different sign to $J_{kj}$, we assign the value $J_{kj} - \eta$ to it instead.

The first feature of the results to note is that in both versions of the PEC model, the power law relationship between the size and frequency of extinctions breaks down and no longer offers a description of the data. There are far more very large extinctions in the

---

[2] the unchanged update rule means that same-sign interactions between agents can exist, but in practice these make up less than 1 per cent of the total number of interactions

PEC system. Figures 1a, b and c plot the relationships between extinction size and frequency in typical solutions of the model with N = 100 for, respectively, version A of the PEC model, version B of the PEC model , and the general version of the model.

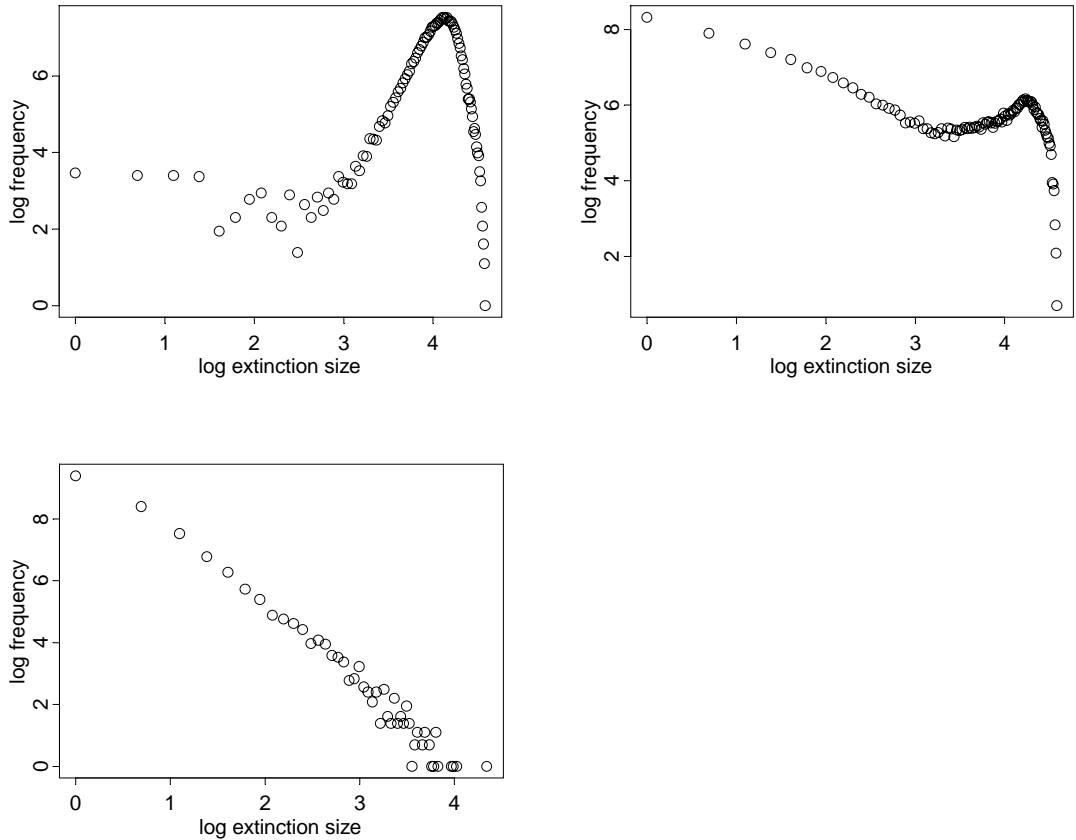

*Figures 1a to c. Plots showing extinction size vs. frequency with logarithmic axes for typical solutions of (a: top left ) PEC model version A (b: top right) PEC model version B. (c: bottom ) general S-M model.*

As a consequence of the high frequency with which large extinctions are observed, the average overall level of fitness of the system remains very low. For example, with 100 agents, the average level of fitness of the model over 500 simulations is only 0.0014 and 0.056 in versions A and B of the PEC model, compared to 0.12 in the general version. Figure 2 shows the average level of fitness in the two versions of the PEC model and in the general model for N = 100       .

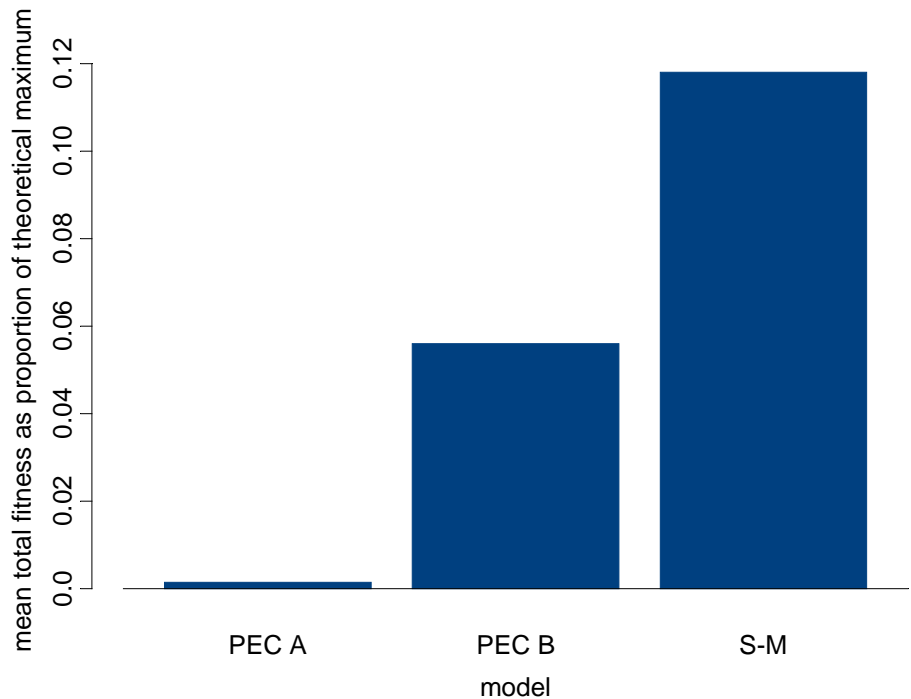

*Figure 2. Bar chart showing the average total fitness (as a proportion of theoretical maximum fitness) for the three models.*

The overall level of fitness of the system is very low in version A of the PEC model. The higher level in version B is accounted for by the pattern of extinctions over time  Figures 3a and b plot the numbers of agents becoming extinct in each period of a typical simulation of the general model and version B of the PEC model.  In model B, extinctions take place in bursts, with relatively long periods of very low extinction levels being interspersed by periods of very high levels.  The average overall fitness of the system over time is raised by the former.

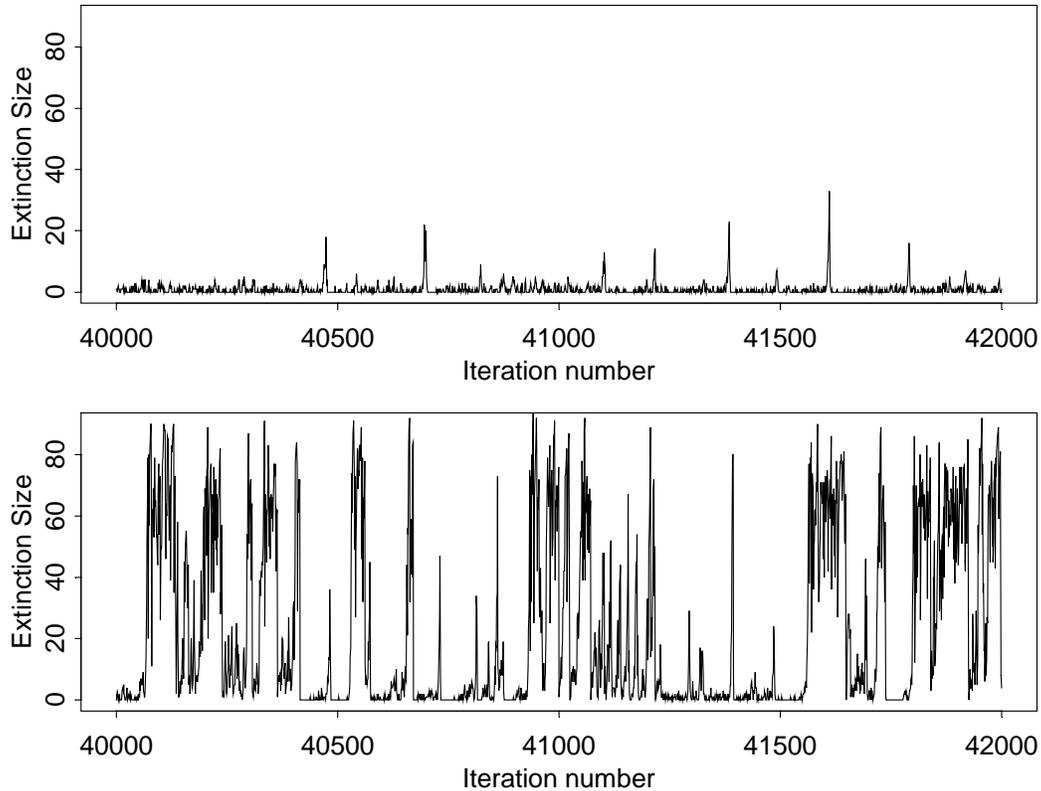

*Figures 3 a and b. Typical extinction size patterns, plotted over 2000 mid-run iterations, for (top a) the S-M model and (bottom b) PEC model B, each with 100 agents.*

Overall, these results suggest that systems in which the only interactions of agents which are permitted are those implied by economic competition, do not generate the type of power law behaviour which is observed in actual economies. Further, a certain amount of collaboration and co-operation between firms in any given industry appears to be necessary for the industry as a whole to acquire viability in terms of overall fitness. The enforcement of strict economic competition between firms by regulatory authorities, particularly in industries in which entry is relatively easy, runs the risk of causing large scale extinctions of companies.

## 5. Conclusion

Biological species interact with each other in ways which either increase or decrease the fitness levels of individual species. The outcome of interactions between firms in the economy on each others' fitness levels can be thought of in the same way, as the impact of the strategy of individual agents on other agents. Indeed, there is evidence of similarity of the empirical power law which relates the size of an extinction to its frequency in the extinction records of both biological species and firms in the economy.

In economic theory, it is regarded as desirable that firms in the same industry should compete with each other, and should not enter into collusive or co-operative behaviour. This is the principle which underlies a great deal of the legal regulation of competition in industries which exists in the West.

In this paper, we summarise the properties of a general model of agent interaction which has been used to account for the empirical relationships observed in the fossil record of the extinctions of biological species. We contrast these with those of a pure economic competition version of the model, in which co-operative or collusive behaviour between agents is not permitted.

The special case of the pure economic competition model has quite different properties from the general model of agent interaction. A power law no longer gives a good account of the relationship between extinction size and frequency. In other words, systems in which the only interactions permitted are those implied by economic competition, do not generate the type of power law behaviour which is observed in actual economies. From a policy perspective, the overall level of fitness is much lower in the pure competition model. This suggests that a certain amount of co-operation between firms is a necessary condition for the survival of an industry.